\documentclass[journal,compsoc]{IEEEtran}
\ifCLASSOPTIONcompsoc
  \usepackage[nocompress]{cite}
\else
  \usepackage{cite}
\fi
\ifCLASSINFOpdf
  \usepackage[pdftex]{graphicx}
\else
\fi
\usepackage{array}
\usepackage[hyphens]{url}
\usepackage[utf8]{inputenc}
\usepackage{csquotes}
\usepackage{dirtytalk}
\newcolumntype{C}[1]{>{\centering\let\newline\\\arraybackslash\hspace{0pt}}m{#1}}

\usepackage[capitalize]{cleveref}

\usepackage{textcomp} %required to use upquote below
\usepackage{listings}
\begin{document}
\title{Towards Automated Performance Bug Identification in Python}

%\subtitle{A case study on a Real Time System}


\author{Sokratis~Tsakiltsidis,
        Andriy~Miranskyy,
        and~Elie~Mazzawi% <-this % stops a space
\IEEEcompsocitemizethanks{\IEEEcompsocthanksitem S. Tsakiltsidis and A. Miranskyy are with the Department of Computer Science, Ryerson University, Toronto, Canada. \protect\\
E-mails: stsakiltsdis@ryerson.ca and avm@ryerson.ca
\IEEEcompsocthanksitem 
E. Mazzawi is with Addictive Mobility, Toronto, Canada}% <-this % stops an unwanted space
}


\maketitle

%%%%%%%%%%%%%%
%% ABSTRACT %%
%%%%%%%%%%%%%%
\begin{abstract}
\textit{Context}: Software performance is a critical non-functional requirement, appearing in many fields such as mission critical applications, financial, and real time systems. In this work we focused on early detection of performance bugs; our software under study was a real time system used in the advertisement~/~marketing domain.

\textit{Goal}: Find a simple and easy to implement solution, predicting performance bugs. 

\textit{Method}: We built several models using four machine learning methods, commonly used for defect prediction: C4.5 Decision Trees, Na\"{\i}ve Bayes, Bayesian Networks, and Logistic Regression.

\textit{Results}: Our empirical results show that a C4.5 model, using lines of code changed, file's age and size as explanatory variables, can be used to predict performance bugs (recall~=~0.73, accuracy~=~0.85, and precision~=~0.96). We show that reducing the number of changes delivered on a commit, can decrease the chance of performance bug injection. 

\textit{Conclusions}: We believe that our approach can help practitioners to eliminate performance bugs early in the development cycle. Our results are also of interest to theoreticians, establishing a link between functional bugs and (non-functional) performance bugs, and explicitly showing that attributes used for prediction of functional bugs can be used for prediction of performance bugs.

\end{abstract}


% \fi
\IEEEpeerreviewmaketitle

%%%%%%%%%%%
%% INTRO %%
%%%%%%%%%%%
\section{Introduction} \label{intro}

There exists large number of fields where performance of software is critical. For example, mission critical applications, financial and real time systems \cite{han2012performance, jin2012understanding, killian2010finding}. 

Based on statistics from the International Advertising Bureau, mobile ad expenses for 2014 reached \$31.9 billion US dollars \cite{iab}, followed by a 20\% growth in 2015. This revenue is entirely managed by real time buying systems.

Real time systems (RTS) are software that subject to time constraints; they have to process information and provide an output within a pre-specified time threshold. A more specific sub-category of RTS is the application under study, which is a real time buying (or bidding) system (RTB), used in the advertisement/marketing domain. The RTB system, also known as Demand Side Platform (DSP), receives an \textit{http} request from the Sell Side Platform (SSP). This request, which is basically an auction being held by the SSP, contain information that is processed by the RTB. The RTB responds back with an advertisement, if some (or all) of the criteria are met. If the RTB system does not respond within a 100 milliseconds time constraint, the opportunity to bid is lost.

Code inefficiencies for this type of software can easily cause reduced throughput, increased latency, and wasted resources. All the above further translate to unreliable software that cannot respond within the desired time threshold (the aforementioned 100ms), leading to financial losses, due to unsuccessful bidding. Also, since performance bugs are not reported as often as functional bugs \cite{jin2012understanding, zaman2011security}, it is easier for them to accumulate and affect the system for a longer time. In this study (in line with \cite{jin2012understanding, killian2010finding, zaman2012qualitative}) we will define \textit{performance bug} as a software  defect,  where  after  relatively  simple  changes  are  applied,  the  performance of  the software  is  significantly  increased  without  affecting functionality.

\textbf{Problem:} Continuous integration practices are gaining momentum and are being used widely, especially by startups \cite{fowler2006continuous}. However, lack of time for sufficient performance testing and immaturity of the development team can cause performance bugs to be injected at all times. The problem in such cases, in contrast with regular bugs that affect the system’s outcome, is that such types of bugs are not easily diagnosed. Even if they are, developers can easily omit fixing/improving them in favor of delivering new features or other more significant tasks, such as fixing security bugs \cite{zaman2011security}. We need a way (preferably a simple and tractable one) to detect performance bugs using information that is either readily available to a developer or can be easily obtained. Static analysis and use of attributes extracted from source code repository are promising candidates.  

Moreover, we want to understand which factors (i.e., attributes associated with committed code) affect injection of performance bugs, so that the management can take corrective actions. 
Therefore, our \textbf{research question} is: \\  

%% RQ 
\emph{How can we detect performance bugs using static code analysis and attributes extracted from source code repository?} \\

\textbf{Approach:}  We tackle the research question by performing a case study; our software under study is a RTB system developed over a period of three years. Our goal was to determine if the commit metrics that we gathered, are associated with the probability of performance bugs to be injected. Approximately 2800 file commits (delivered over this three year period) were processed in order to extract a variety of attributes and metrics (analyzed in Section \ref{res_dataset}) for every file that was changed during development.

These attributes and metrics were categorized into three groups: a) Project attributes, b) Activity attributes, and c) Experience attributes. Then the data have been preprocessed (see Section \ref{data_prep} for details). We tried several approaches both in the manner of attributes selected as well as the classification algorithms that were used, in order to achieve the best outcome. We also quantified the contribution of each attribute on the efficiency of each model.

\textbf{Contributions:} Our main contributions from this work are: a) explicitly establishing that static code analysis and metrics can be used as predictors of performance bugs, similar to functional bugs, b) understanding the differences between classification algorithms that are commonly used to predict bugs, c) understanding the impact of each attribute on the models' efficiency. 

Performance bugs are often caught while timing reference workloads. However this is typically done late in the development cycle. Our approach is complementary; it enables early detection (and thus early elimination) of the performance bugs, since prediction models operate on the source code and do not require code execution.

\textbf{Impact:} What we present in this paper is a starting point for practitioners seeking performance optimization and an efficient way of predicting performance bugs. A successful detection will have a significant impact, since in general, performance bugs need much more time to get fixed than non-performance (functional) bugs \cite{zaman2011security, zaman2012qualitative}.

\textbf{Structure of this work:} We provide a review of related work that has been conducted in this area, and identify the knowledge gap in Section \ref{rel_work}. Next, in Section \ref{methodology}, we present the methodology that we followed and the design of our study. Results of the study are given in Section \ref{results}. Threats to validity are identified in Section \ref{threats}. Finally, Section \ref{conclusion} summarizes our thoughts and outcomes, and specifies future research potentials.

%%%%%%%%%%%%%%%%
% RELATED WORK %
%%%%%%%%%%%%%%%%
\section{Related Work} \label{rel_work}
There are many studies in the area of defect prediction (see Hall et al. \cite{hall2012systematic} for review of 208 defect prediction models)  and quality assurance of software, since this is one of the most significant endeavors during a product’s life cycle. Reduced number of bugs assures better quality, thereby improving the product being delivered and allowing for better resource allocation \cite{fenton1999critique, hall2012systematic}.

There are several categories of tools and methods used to detect and predict defects, with the aim of delivering better software \cite{orso2014software}. However, static bug finders and statistical models \cite{rahman2014comparing} have become the two most prominent categories of defect prediction. The defects can be further categorized to create prediction models in order to distribute the resolution resources more efficiently \cite{caglayan2015predicting, caglayan2010usage, kan_metrics_2002}.
At the same time, efficiency, density, and principally the technical debt caused by defects, play critical roles as well \cite{vetro2012using}.

Failures after the release of a product have also momentous effects, especially for large-scale commercial distributions \cite{jin2012understanding}. Even though developers dedicate large amounts of effort and time to testing, they can never be sure that the system is absolutely reliable.

\emph{Performance}

Even though there are many in-depth studies trying to determine the best way of predicting functional bugs, there is not much work done in prediction of bugs causing performance degradation. Below we give an overview of research papers studying performance bugs.

Performance bugs are harder to expose during testing phase, because they do not cause fatal symptoms and do not affect the overall outcome \cite{molyneaux2014art}.  Furthermore, it is difficult to find the root cause of performance bugs (in comparison with other types of bugs); they also need more time to get resolved \cite{zaman2011security}.

The work that is closest to ours is by Jin et al. \cite{jin2012understanding}, which studied a set of 109 real-world performance bugs. They studied the bugs’ lifetime from inception to fix, their root causes and introduction mechanisms, in order to create rule-based detectors. We used a comparable approach to create detectors (referred to as "patterns" in our study) with the aim to identify similar performance bugs. They studied software written in Java, C, C++, and JavaScript; ours was written in Python\footnote{Programming language does affect performance of the models, as shown by \cite{weyuker_2010}.}. Moreover, we focused on finding an efficient method to automatically detect performance bugs based on data extracted through the use of patterns (197 real-world performance bugs in our case). Summarizing, Jin et al. \cite{jin2012understanding} focused on analyzing characteristics of performance bugs, while in this study, our focus is on understanding the contribution of each source code attribute, as extracted from the source code repository, to the predicting power of the several machine learning algorithms that we used. Lastly, our general approach and the methodology for creating the prediction model were meant to be easily reproducible in the future. Therefore, our work is complementary.

Automated detection of performance bugs have been implemented in the past \cite{foo2015, killian2010finding}. However, the authors used dynamic analysis tools, namely execution traces and historical performance data, to detect slowly executing code. In addition, dynamic analysis tools often require dedicated testing environment to get accurate performance readings. We, on the other hand, are leveraging code attributes extracted from source code repository and automatically detecting performance bugs before the code executes (and reaches production environment). Thus, our work is complementary.  

Static analysis tools can also eliminate performance bugs~\cite{nistor2015}. However, this approach requires knowledge of pattern template, while ours does not (as we omit pattern type info in our models), hence the complementarity of our work.

There exist formal and thorough coding standards for RTS. However, the standards are, typically, language specific (e.g., C~\cite{misra-c} and C++~\cite{misra-cpp}). To the best of our knowledge, no thorough and formal RTS coding standard exists for Python.

Finally, since in our study we relate performance with the real-time nature of the system, one could argue that Python programming language might not be appropriate for such a case. Although this is a problem that falls out of the scope of this study, we understand its importance and validity. Python programming language (with a proper runtime \cite{paypal}) is used in well-known large-scale RTS, such as eBay, PayPal, and YouTube \cite{paypal, python_quotes}. The benefits of the language are best summarized by Cuong Do, Software Architect of YouTube: ``Python is fast enough for our site and allows us to produce maintainable features in record times, with a minimum of developers'' \cite{ python_quotes }.  

%%%%%%%%%%%%%%%
% METHODOLOGY %
%%%%%%%%%%%%%%%
\section{Methodology} \label{methodology}
In this section we discuss methodology of our case study. The section is structured as follows. We describe our software under study in Section \ref{e_study}. The dataset (extracted from the software under study) and its preprocessing are depicted in Sections \ref{res_dataset} and \ref{data_prep}, accordingly. Finally, we discuss building of prediction models in Section \ref{model_building}.

\subsection{Software Under Study} \label{e_study}
The software that we worked on is a RTB application, written in Python behind an asynchronous web framework. The system has been developed from scratch during the last three years. Although the history of the code might sound short, due to the continuous integration practices, the commit count is 2773 and that is only for the production (master) branch. Therefore, we had sufficient amount of information to work with.

The RTB application receives as input \textit{http} requests and responds back within a time constraint (typically, 100 milliseconds), that is set by the SSPs, which the RTB is connected with. The load of the system can grow tremendously, based on various factors, e.g., incoming request rate and current advertisement inventory.

It is the nature of the system that makes it vulnerable to small fluctuations of the response time, and that is where performance becomes critical. Although in common systems availability is measured by subtracting downtime, in our case, we have to append the total of timed-out sessions in order to calculate downtime. During a timeout, the system’s functionality is not affected. However, the strict time threshold is not allowing the system to complete a computational cycle successfully.

\subsubsection{Infrastructure}
Since performance is usually related to workload, we provide some information regarding the infrastructure associated with the software under study.
The cloud cluster undertaking the load consists of 22 virtual machines, each containing 32~vCPUs\footnote{Intel Xeon E5-2680 v2 (Ivy Bridge) Processors.} and 60~GB of RAM, each one of them capable of processing more than 3,000 requests per second. The cluster processes, on average, 10 billion transactions per day (going up to 30 billion transactions on a busy day). However, it is worth mentioning that the amount of machines deployed is dynamic and depends on many factors, such as high peak time periods of the year or the number of countries that the current ads are targeting (the larger the number of countries -- the higher the load). In such cases more computational power is added in order to accommodate the load.

\subsubsection{Spot the Bottlenecks}
As a first step of our methodology, we had to identify the bottlenecks of the system. This was achieved with the help of dynamic program analysis tools while reproducing conditions that happen in production environment and, in some cases, executing the analysis on the production environment itself. Although dynamic analysis tools -- the well-known \textit{profilers}, have been improved extensively and can provide accurate results, they cannot identify the performance bugs themselves \cite{jovic2011catch}. However, they can help in preliminary analysis and assist in easy verification of the experimental findings.

By identifying the bottlenecks, we were able to find alternative ways that optimize the performance of the system. Most of the times the changes that had to be applied were a few lines of code, which resulted in significant performance improvements. These findings were categorized and classified as \textit{patterns}.

\subsubsection{Patterns} \label{patterns}
We define patterns as coding practices/styles followed (coincidentally or deliberately) by individual developers.  These patterns introduced a logic that affected the performance, but not the overall correctness of the algorithm/system. We identified eleven patterns, described in Appendix~\ref{app:patterns}, causing performance degradation (PD). Identification was performed by detecting bottlenecks (i.e. code blocks causing PD) using profilers, followed by manual code inspection. All patterns always lead to PD, which we validated on four platforms and two Python versions. We fixed 197 instances/defects, based on these patterns; fixing patterns leads to 20-99\% improvement for a given code block; 99\% improvement was achieved by fixing pattern described in Appendix~\ref{p:2}.

These eleven unique patterns were generalized and used in scripts that we built, so that we could detect multiple areas of application. This type of static bug finders were used in the past for similar purposes \cite{jin2012understanding, livshits2005dynamine}; however, in this study they were mainly used to get an enriched dataset.

The practice of pattern identification was used before by Jin et al. \cite{jin2012understanding} (rule-based detectors) in order to achieve similar results, i.e., detect multiple areas that were ``affected'' by the same generalized model. However, their patterns were mainly connected to outdated or error prone API's and, in most of the cases, the bugs were product-specific. Contrariwise, the patterns that we identified are more general. Essentially, they can be re-used on any software written in the same programming language (Python) and are more related to the development techniques utilized by the software developers.\\

\emph{Benchmarking Patterns}\\

In terms of benchmarking and calculating the performance improvement of each pattern, we thoroughly tested the impact of each one of them. The performance improvements per pattern ranged between 20\% and 99\%.  

Additionally, and in order to validate that a pattern is not endemic to a particular environment (e.g., Operating System, Hardware Platform, or Programming Language version) we tested them in multiple environments and observed that they lead to similar performance degradation in all environments.

\subsection{Resulting dataset} \label{res_dataset}
Having the patterns as a stepping-stone (and the static bug finder scripts using them), we could move forward and extract a dataset with the help of the  version control system. It is worth mentioning that the dataset that we extracted and utilized, contains information on the performance bug injection points. Such information is difficult to extract from a bug tracking system \cite{serrano2005bugzilla}, even when dealing with more general types of bugs.

Therefore, our dataset included performance bug injection commits and was further split by file. As a result we had a thorough dataset including the following attributes\footnote{This is not an exhaustive list of attributes that can be used to predict injection of a defect~\cite{ccauglayan2014factors, li2009analysis, caglayan2012factors}. However, we focused on the attributes that can be easily extracted by practitioners to simplify institutionalization of the prediction models. }: \\

\emph{Project attributes}
\renewcommand{\labelenumi}{\roman{enumi}}
\begin{enumerate}

\item{\textit{\textbf{Type:}} Indicates either if the entry injected\footnote{We manually traced injection point for each performance bug using history information obtained from the source code management system.} a performance bug into the code or not (not a defect at all).} 
\item{\textit{\textbf{Commit ID:}} Unique identifier of the commit. 

\textit{Total number of commits is 2773.}}
\item{\textit{\textbf{Owner:}} The user that submitted the commit.

\textit{13 unique owners under study.}}
\item{\textit{\textbf{File name:}} The file that was modified. Since each commit can contain several files, we split the entries to unique 2-tuples of commit id and file name (henceforth called tuple). In essence, the tuple serves as a unique identifier for a record in our dataset.

\textit{338 unique files, 4623 unique tuples (commit id and file name).}}
\end{enumerate}

\emph{Activity attributes}

\begin{enumerate}
\setcounter{enumi}{4}

\item{\textit{\textbf{Lines added:}} The number of lines that were added (or modified) for the corresponding file. Note that one modified line will be counted as one line removed and one line added (a-la diff). }
\item{\textit{\textbf{Lines removed:}} The number of lines that were removed (or modified) for the corresponding file.}
\item{\textit{\textbf{File age:}} Age of the file, computed as `modification date -- file creation date'.}
\item{\textit{\textbf{File size:}} Although two separate values were extracted to represent file size, we treat them as one attribute:}
\begin{itemize}
\item{\textit{Source lines of code.}
Lines of code without comments or blank lines.}
\item{\textit{Comment lines.}
Lines of code representing the developer's comments.}
\end{itemize}
\end{enumerate}

\emph{Experience attributes}

\begin{enumerate}
\setcounter{enumi}{8}

\item{\textit{\textbf{Maturity/Expertise of the owner:}} The amount of time an owner spent working on the product, computed as `modification date -- first commit date of the owner'.}
\item{\textit{\textbf{Time since last commit:}} How much time elapsed since the last commit (in any of the files).}
\end{enumerate}

\begin{figure*}[!ht]
\centering
\includegraphics[width=\linewidth]{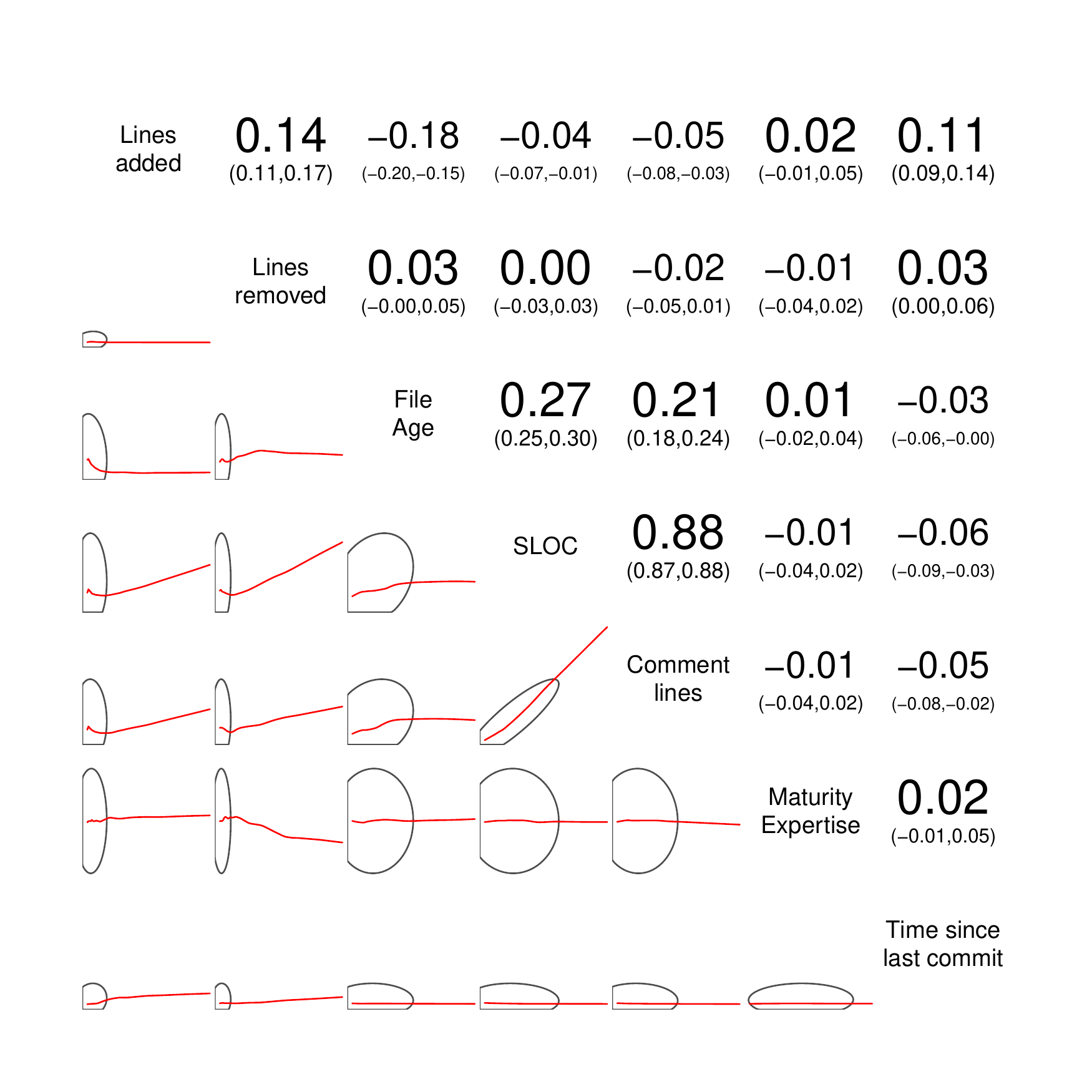} 
\caption{Correlogram visualizing correlation matrix \cite{corrgram} for numeric attributes. Diagonal shows attribute name. Lower triangle region shows confidence ellipse and smoothed line. Upper triangle region shows Pearson’s correlation coefficient $\rho$ and, in brackets, confidence interval for $\rho$. }
\label{fig:cor}
\end{figure*}

Descriptive statistics for all the numerical attributes is shown in Table \ref{tab:attrs}. Correlation between the numerical attributes is given in Figure \ref{fig:cor}. As we can see, none of the variables, except\footnote{Which is typical for a software product.} `source lines of code' (SLOC) and `comment lines', are strongly correlated.

\begin{table*}[ht]
\centering
\caption{Descriptive statistics of numeric attributes.}

\begin{tabular}{|l|c|c|c|c|}
\hline 
\textbf{Attribute} & \textbf{Range} & \textbf{Mean} & \textbf{Median} \\ 
\hline \hline
\rule{0pt}{2ex}
Lines added & 0 - 642 & 19.74 & 3 \\ 
\hline 
\rule{0pt}{2ex}
Lines removed & 0 - 995 & 13.21 & 2  \\ 
\hline
\rule{0pt}{2ex}
File age (days) & 0 - 803 & 139.2 & 59.97  \\ 
\hline
\rule{0pt}{2ex}
SLOC & 0 - 1109 & 254.2 & 150  \\ 
\hline
\rule{0pt}{2ex}
Comment lines & 0 - 419 & 71.07 & 30  \\ 
\hline
\rule{0pt}{2ex}
Maturity/Expertise (days) & 0 - 997 & 416.9 & 435  \\ 
\hline
\rule{0pt}{2ex}
Time since last commit (hours) & 0 - 266.7 & 10.71 & 1.26 \\
& ($\approx 11$ days) & & \\
\hline
\end{tabular}

\label{tab:attrs}
\end{table*}

\subsection{Dataset Preprocessing} \label{data_prep}

\subsubsection{Cleansing}
We eliminated all files that did not go into production code base, namely: readme files, testing scripts, and help files.

Additionally, we removed a minor 0.2\% (9 out of 4623 unique tuples) of ``commit id -- file name'' records related to source code files. These records were outliers, extreme cases. For example, we excluded source files that were moved or removed. To be more specific, the  version control system by default identifies directory changes/refactorings as complete removals of the files themselves. Therefore, whenever a file is moved one or more levels up or down in the directory structure, we noticed abnormal numbers of lines added and/or removed. In some of theses cases (and especially in directories including large files) more than 10,000 lines were added or removed on a single commit. The cleansing described above, resulted in a more precise model creation, mainly because of the removal of entries that will not be seen on future commits \cite{gray2011misuse}.

\subsubsection{Re-balancing}\label{rebalancing}
After the initial cleansing we had our dataset in its final form. However, and as a result of the large number of commits that we have under study, it was expected that the non-defective entries would prevail over the defect set. In fact, we had 95.5\% (4426) of non-defective entries and 4.5\% (197) of defective ones. This type of imbalance is common, especially when dealing with real world problems \cite{he2009learning}. Therefore, in order to avoid biased results towards the majority class, we had to choose a re-balancing technique.

There exists several re-balancing methods, since this ubiquitous problem of imbalanced data is faced quite often. To overcome this obstacle and achieve valid results, researchers use over and under-sampling techniques, either in the minority or majority classes. While reviewing these methods and techniques that have been used towards this goal, we decided to use random over-sampling for our training dataset. Against the numerous (and in some cases combined and complex) methodologies, as studied by Batista et al. \cite{batista2004study}, random over-sampling provides competitive results.

\subsubsection{Overcoming Premature Over-fitting} \label{overcome_pof}

Given high imbalance of our data, we expected that models trained on rebalanced datasets would be prone to over-fitting (as seen by \cite{mease_2007, he2009learning}). This was confirmed during preliminary model building (discussed in Section~\ref{model_building}): we observed over-fitting of the models, no matter the classification method. 

As described by Tan et al. \cite{tanonline}, over-fitting may occur even when one utilizes cross-validation, especially when dealing with extremely imbalanced datasets. Even after applying the re-balancing methods (described in Section~\ref{rebalancing}), the repetition of the entries might still result in false high precision \cite{tanonline}. 

To overcome this issue, we used a combination of techniques, similar to the one that was introduced by Mease et al. \cite{mease_2007} and called ``JOUS-Boost''. They added independently and identically distributed noise to the minority class for which random oversampling created replicates. This approach was shown to be superior in dealing with the oversampling problem \cite{he2009learning}.  

In our case, we decided to generate a ``jittered'' train set, after we split our data to 70\% train - 30\% validation. We added noise to the minority entries of the 70\% part, by randomly increasing or decreasing the values by 0 to 2\%. Formally, a variable $x$ becomes $x \times U(0.98, 1.02)$, where $U(0.98, 1.02)$ is a random value drawn from the uniform distribution with minimum value of $0.98$ and maximum value of $1.02$. This transformation could only apply to numerical attributes (listed in Table \ref{tab:attrs}). We do not add noise to the validation set, to mimic the actual defect prediction process; when a developer passes information about a new commit to the prediction `oracle' to predict if this commit contains a performance bug or not. Results of this experiment are discussed in the next section.

\subsection{Building the Models}\label{model_building}

\subsubsection{Classification Algorithms}
After re-balancing the dataset, it is time to move forward and start building the desired models. The first question that arises is: which classification algorithms should be used for building the models? Based on the literature and existing empirical work \cite{moser2008comparative} we decided to experiment with four popular machine learning methods (using Weka \cite{weka} and R \cite{R}). The algorithms chosen are: a) C4.5 decision tree (Weka's J48 class \cite{weka}), b) Na\"{\i}ve Bayes, c) Bayesian networks, and d) Logistic regression.

\subsubsection{Training and Testing the Models} \label{training_models}
We trained and tested the model on the 70\% of the data using 10-fold cross validation; we then validated the model on the remaining 30\% of the data (see Section \ref{overcome_pof} for description of data subsets). We used this approach to generate more precise efficiency metrics for the models. The details of the efficiency metrics --- namely, accuracy, precision, and recall --- are given below. 

 \textbf{Accuracy:} accuracy  reflects the percentage of correctly classified performance bugs to the total number of performance  bugs. It is computed as follows \cite{witten2005data}:

\begin{equation}
    ACC  = \frac{(TP+TN)}{(TP+TN+FN+FP)},
    \label{eq:1}
\end{equation}
where $TP$ is the number of true positive results, $FP$ is the number of false positive results, $TN$ is is the number of true negative results, and $FN$ is the number of false negative results.

\textbf{Precision:} precision, or Positive Predictive Value is the fraction of retrieved instances that are relevant, and it is calculated as \cite{witten2005data}:

\begin{equation}
    PPV  = \frac{TP}{(TP+FP)}.
    \label{eq:2}
\end{equation}

\textbf{Recall :}  recall, also known as sensitivity or True Positive Rate, is the fraction of relevant instances that are retrieved, and it is calculated as \cite{witten2005data}:

\begin{equation}
    TPR  = \frac{TP}{(TP+FN)}.
    \label{eq:3}
\end{equation}

All three metrics range between 0 and 1; the closer the value to 1 -- the better.

\begin{table*}[ht]
\centering
\caption{Experiment metrics: JOUS-Boost. The models are cumulative: the first model describes the model using one explanatory variable: lines added; the second model – two variables: lines added and lines removed;  the third one – three variables: lines added, lines removed, and age; and so on. Size consists of two separate attributes, as described in Section \ref{res_dataset}.}
\label{table:jit}
\begin{tabular}{|c|l|c||C{2cm}|c|c|C{2cm}|}
  \hline 
  %\multicolumn{2}{| c ||}{}
   & & & & & & \textbf{Logistic} \\
  %\multicolumn{2}{| c ||}{}
  \textbf{Row \#} & \multicolumn{1}{c|}{\textbf{Variable}} & \textbf{Metric} & \textbf{C4.5} & \textbf{Na\"{\i}ve Bayes} & \textbf{Bayesian Nets} &     \textbf{Regression} \\
  \hline
  \hline
    \rule{0pt}{2ex}
     & +Lines added & TPR: & 0.65 & 0.52 & 0.68 & 0.67 \\
    1 & \textit{jittered} & ACC: & 0.81 & 0.74 & 0.82 & 0.79 \\
     & & PPV: & 0.95 & 0.93 & 0.95 & 0.89 \\
  \hline
    \rule{0pt}{2ex}
     & +Lines removed & TPR: & 0.67 & 0.52 & 0.67 & 0.67 \\
    2 & \textit{jittered} & ACC: & 0.82 & 0.73 & 0.82 & 0.79 \\
     & & PPV: & 0.95 & 0.91 & 0.95 & 0.89 \\
  \hline
    \rule{0pt}{2ex}
     & +Age & TPR: & 0.70 & 0.52 & 0.67 & 0.68 \\
    3 & \textit{jittered} & ACC: & 0.84 & 0.73 & 0.82 & 0.80 \\
     & & PPV: & 0.96 & 0.91 & 0.95 & 0.89 \\
  \hline
    \rule{0pt}{2ex}
     & +Size & TPR: & 0.73 & 0.83 & 0.58 & 0.70 \\ 
    4 & (SLOC \& Comments) & ACC: & 0.85 & 0.73 & 0.79 & 0.80 \\
     & \textit{both jittered} & PPV: & 0.96 & 0.69 & 0.98 & 0.88 \\
  \hline
    \rule{0pt}{2ex}
     & +Experience & TPR: & 0.73 & 0.83 & 0.58 & 0.70 \\ 
    5 & \textit{jittered} & ACC: & 0.85 & 0.74 & 0.79 & 0.80 \\
     & & PPV: & 0.96 & 0.70 & 0.99 & 0.87 \\
  \hline
    \rule{0pt}{2ex}
     & +Time since last commit & TPR: & 0.72 & 0.88 & 0.52 & 0.70 \\ 
    6 & \textit{jittered} & ACC: & 0.84 & 0.72 & 0.76 & 0.80 \\
     & & PPV: & 0.96 & 0.67 & 0.99 & 0.88 \\
  \hline
    \rule{0pt}{2ex}
     & +Owner & TPR: & 0.68 & 0.90 & 0.52 & 0.77 \\ 
    7 & & ACC: & 0.83 & 0.73 & 0.76 & 0.83 \\
     & & PPV: & 0.97 & 0.68 & 0.99 & 0.87 \\
  \hline
    \rule{0pt}{2ex}
     & +File name & TPR: & 0.83 & 0.88 & 0.57 & 0.90 \\ 
    8 & & ACC: & 0.90 & 0.85 & 0.78 & 0.91 \\
     & & PPV: & 0.96 & 0.84 & 0.99 & 0.92 \\
  \hline
\end{tabular}

\end{table*}

%%%%%%%%%%%
% RESULTS %
%%%%%%%%%%%
\section{Results} \label{results}
The baseline of the attributes used for building each model, is the \textit{lines added} to each file on a single commit. Moving forward, new attributes are added (cumulatively) and their impact on the efficiency can be determined by the change of the metrics.  

The performance of the models is given in Table~\ref{table:jit} and Figures~2-5. Bayesian Nets model yields the best results in the case of a single explanatory variable (\textit{lines added}) with $TPR = 0.68$, $ACC = 0.82$, and $PPV = 0.95$. However, as more explanatory variables are added to the model, C4.5 becomes the top performer. If we exclude the models that use \textit{file name} as the explanatory variable, the best model is based on C4.5 that is using \textit{lines added}, \textit{lines removed}, \textit{age}, and \textit{size} attributes as explanatory variable: $TPR = 0.73$, $ACC = 0.85$, and $PPV = 0.96$. Na\"{\i}ve Bayes and Logistic Regression models, in most cases, take second and third place in the performance competition (surpassing Bayesian Nets). 

\begin{figure}[ht]
    \centering
    \includegraphics[width=\columnwidth]{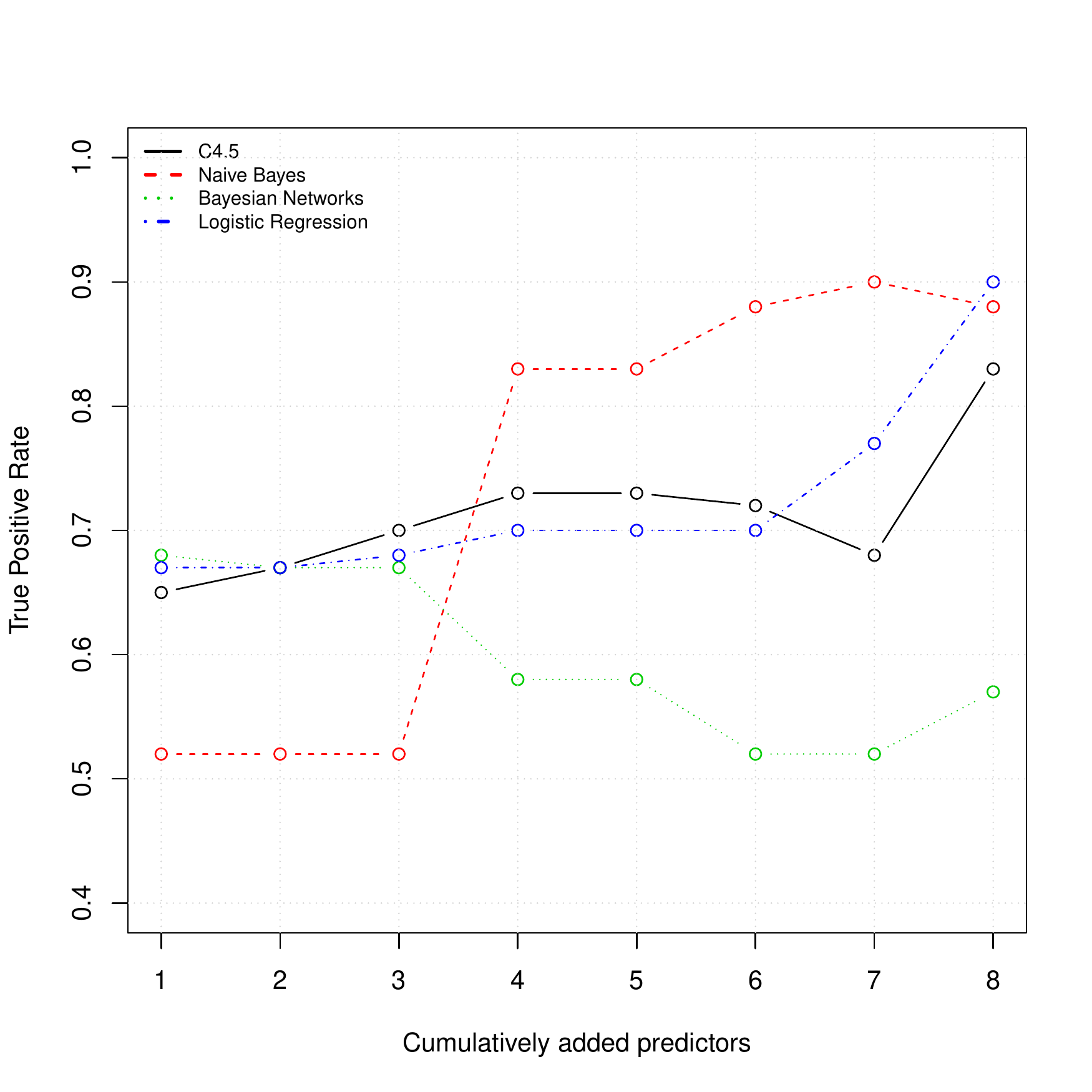}
    \caption{Models' efficiency in term of Recall (True Positive Rate) for cumulatively added predictor variables. The labels of the $x$-axis correspond to the values in the `Row~\#' column of Table~\ref{table:jit}.}
    \label{fig:tpr}
\end{figure}

\begin{figure}[ht]
    \centering
    \includegraphics[width=\columnwidth]{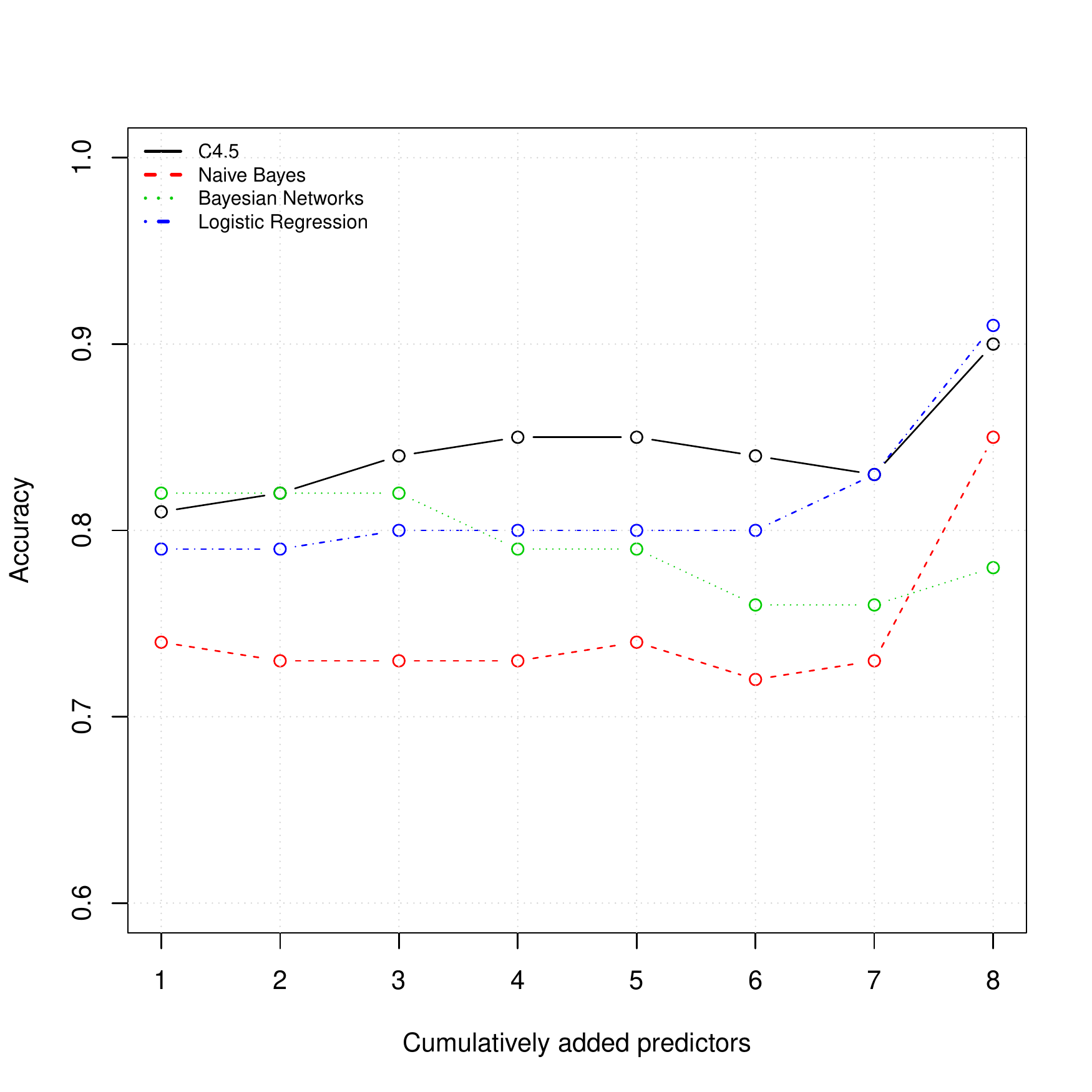}
    \caption{Models' efficiency in term of Accuracy for cumulatively added predictor variables. The labels of the $x$-axis correspond to the values in the `Row~\#' column of Table~\ref{table:jit}.}
    \label{fig:acc}
\end{figure}

\begin{figure}[ht]
    \centering
    \includegraphics[width=\columnwidth]{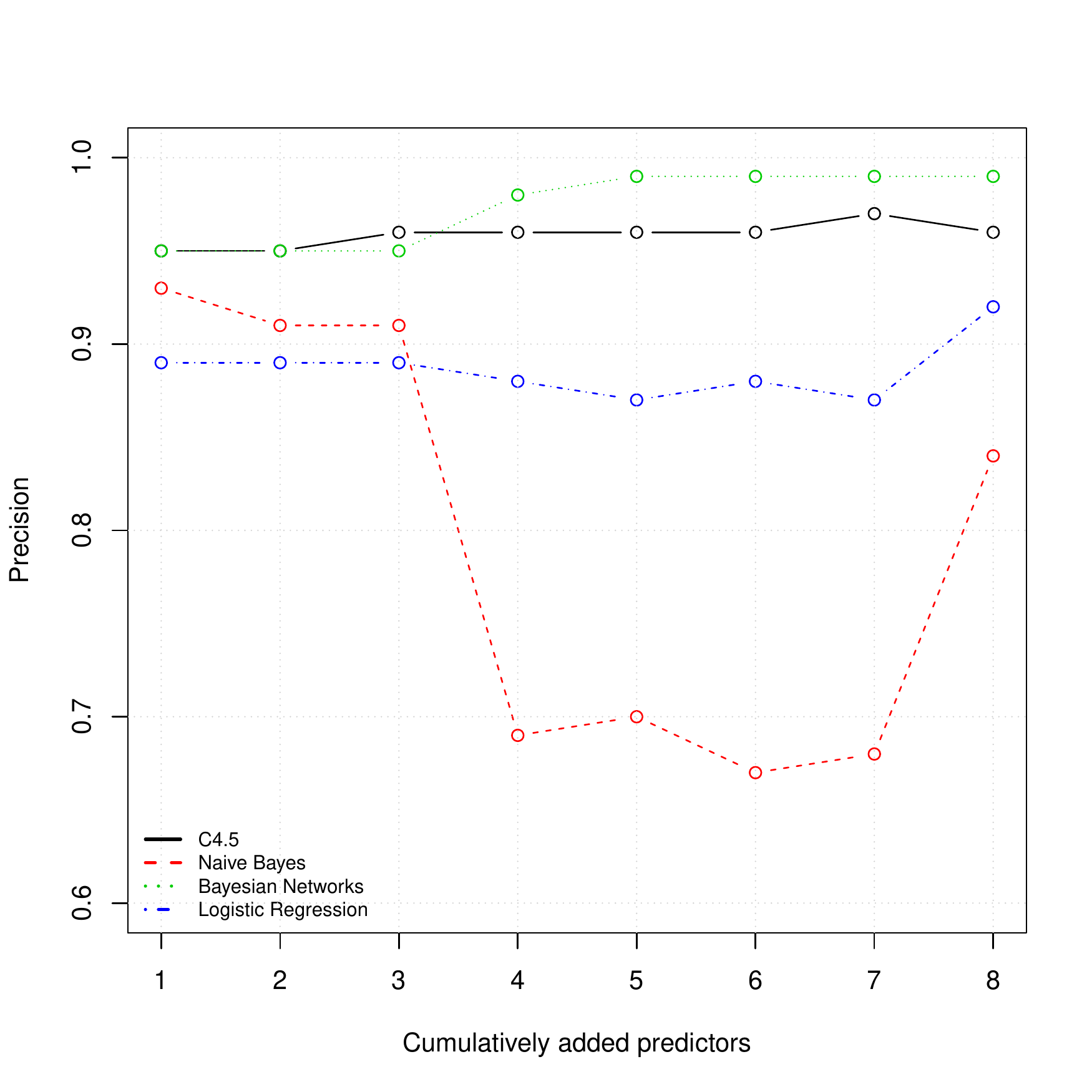}
    \caption{Models' efficiency in term of Precision for cumulatively added predictor variables. The labels of the $x$-axis correspond to the values in the `Row~\#' column of Table~\ref{table:jit}.}
    \label{fig:ppv}
\end{figure}

We will return to the models using \textit{file name} at the end of this section. Let us now analyse which factors affect injection of performance bugs, by analyzing the Logistic Regression (LR) models (because they are the most tractable and reasonably powerful) so that the management can take corrective actions. All the attributes in the LR models are statistically significant: $p$-values $< 0.01$. Recall and Accuracy of the LR models increase as illustrated by positive slopes in Figures~\ref{fig:tpr} and~\ref{fig:acc}, respectively. Precision of the LR models decreases with the addition of the attributes, with the exception of the \textit{file name} attribute (which we will discuss at the end of this section in details), as shown in Figure~\ref{fig:ppv}.

As mentioned above, \textit{lines added} attribute has the highest predictive power. Peculiarly, addition of \textit{lines removed} to the model does not improve its predictive power significantly. This can be explained by the fact that bugs are mostly injected in new code rather than in modified\footnote{ A modified line is counted as one line removed and one line added.} or removed one. The sign of \textit{lines added} regression parameter is positive. This implies that in order to reduce the number of injected performance bugs, the management has to enforce reduction of the amount of changes delivered on a single code commit. This aligns with the findings that the data on the number of lines of code changed provides most of the information needed to predict functional bugs \cite{misirli2011different}. In essence, the more lines you change\footnote{In our case, we can narrow the type of change to code addition.}, the higher the probability that you will inject a bug. Our findings show that this statement is true not only for general functional bugs \cite{misirli2011different}, but also for performance bugs. 

The attributes \textit{age} and \textit{size} improve performance of the models, but not significantly: e.g., compare $TPR = 0.67$ with $TPR = 0.70$. The sign of \textit{age} regression parameter is positive. From the management perspective, this suggests that a performance bug has higher probability of injections, as the software gets older, which is consistent with Lehman's laws of software evolution \cite{Lehman1980}. The signs of \textit{size} attributes \textit{SLOC} and \textit{comments} are negative, suggesting that larger files may have lower chances of injecting performance defect. However, given the low improvement in predictive power, the model may be capturing correlation rather than causality.

Merit of \textit{experience}, \textit{time since last commit}, and \textit{owner} is also arguable. They either do not enhance the predictive power of the model significantly or, in some cases, reduce it, as shown in Table~\ref{table:jit} and Figures~2-4. From the management perspective, this suggests that all developers (independent of their experience with the code base) have injected performance bugs in the past. 

Inclusion of the \textit{file name} improves the performance: the best model is based on Logistic Regression with $TPR = 0.90$, $ACC = 0.91$, and $PPV = 0.92$. This is not surprising, given that our performance bugs reside in 9\% (32 out of 338) of the files. It implies that some files are more prone to performance bugs than others (as seen in other projects \cite{li2009analysis}). The management can focus quality assurance efforts on these 9\% of the files, e.g., making sure that code commits to these files are peer reviewed before merging them into the production code base. 

However, if we reuse our model on a new project, inclusion of the file name attribute into the model may not be practical, since, at the beginning, information about files and new bugs will be limited~\cite{fukushima2014esj}. In this case we can resort to the general C4.5 based model mentioned above: it has lower recall and accuracy, but higher precision.

\begin{figure}[ht]
    \centering
    \includegraphics[width=\columnwidth]{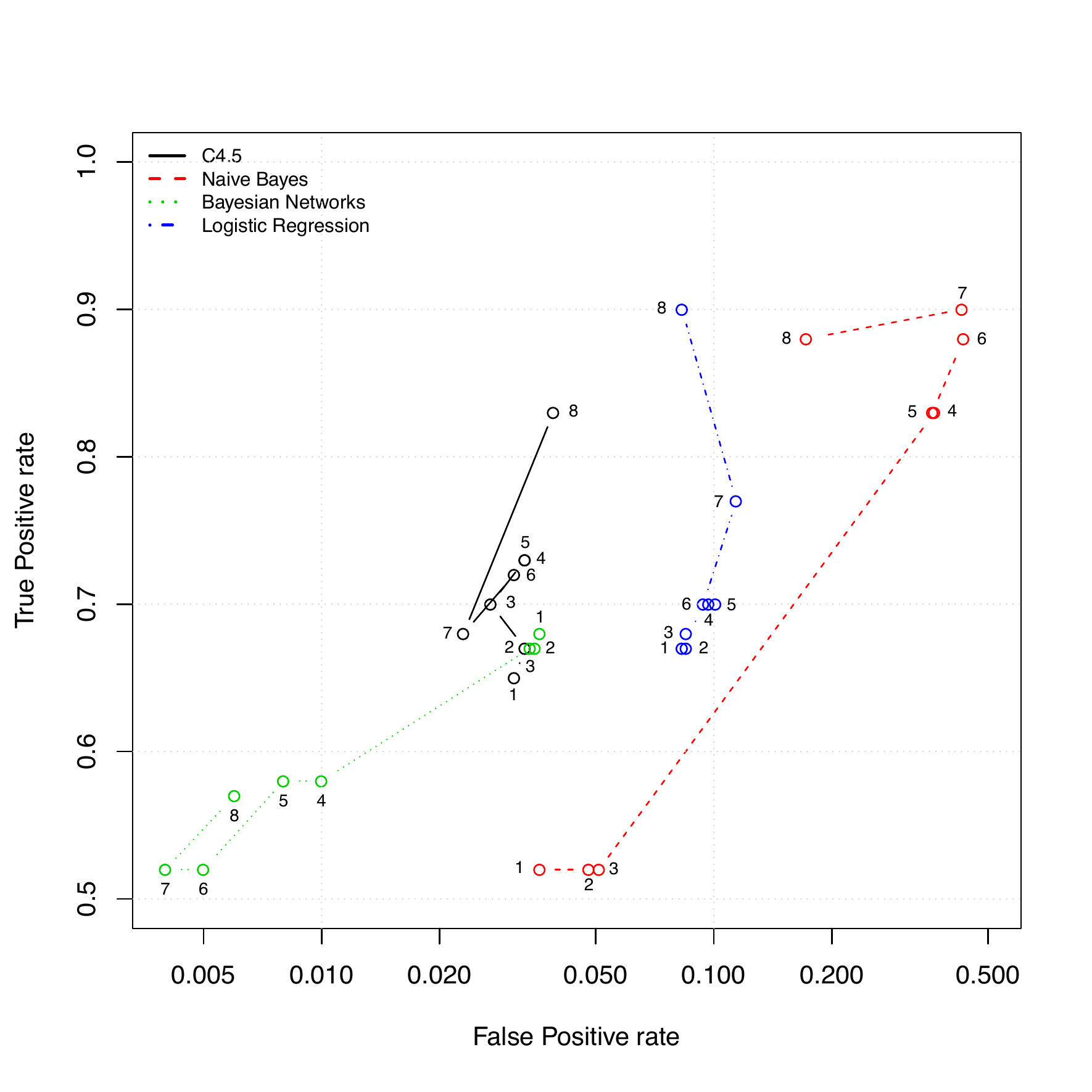}
    \caption{Models' efficiency in term of True Positive Rate vs. False Positive Rate. The points on the graph represent models with different number of cumulatively added predictor variables. Points' labels correspond to the numbers given in the `Row~\#' column of Table~\ref{table:jit}. 
    }
    \label{fig:tp_vs_fp}
\end{figure}

In the final model selection, comparison of True Positive and False Positive rates, shown in Figure~\ref{fig:tp_vs_fp}, might be of help. In this graph\footnote{This graph similar in nature to the receiver operating characteristic (ROC) curve \cite{bradley1997use}.}, we can measure the performance based on the True Positive rate, taking under consideration the drawback of the False Positive rate identification of the models. Since each case and application should be treated differently, we cannot explicitly propose a single model as the best performing one. We will discuss our final selection in Section~\ref{conclusion}.

%%%%%%%%%%%%%%%%%%%%%%%
% THREATS TO VALIDITY %
%%%%%%%%%%%%%%%%%%%%%%%
\section{Threats to Validity} \label{threats}
A number of tests are used to determine the quality of case studies. In this section we discuss four core tests: internal, construct, statistical, and external validity \cite{yin2013case}.

\textbf{Internal:} One of the authors of the study is affiliated with the company. In order to avoid researcher bias, we derived and followed strict automated processes for data extraction and processing. The results of extraction and processing were cross-validated by a researcher not affiliated with the company. 
One of the most critical internal parts of our study is the patterns identification and validation, as explained in Section \ref{patterns}. Validating, benchmarking and leveraging the patterns was one of the most important parts of our study. We tested the patterns in multiple environments (by varying Operating System, Hardware Platform, and Programming Language version) and observed that the patterns lead to software slowdown in all environments.  

Additionally and as already discussed in this paper, the real-time constraints that are externally set for the system under study, as well as the programming language that is used, might lead to bad interpretation of its real-time dynamic. However, this study begun and was applied on an existing, already developed and fully functioning software.

\textbf{Construct:} We construct our dataset based on the data collected from the source code management system. The system does not capture all events and activities happening in the organization. However, based on our results, even this incomplete set of data has strong predictive power.

\textbf{Statistical:} To prevent over-fitting of the models, we utilize 10-fold cross validation for model creation and the JOUS-Boost \cite{he2009learning} approach.

\textbf{External:} Generalizing our findings from a single-case design study to all situations is obviously not possible. However, this design is based on the rationale of the critical case \cite{yin2013case} of a Real Time System. Our results should be transferable to other researchers with well-designed and controlled experiments. In addition, the analysis can be replicated on open source software.

\section{Conclusion}\label{conclusion}
In this work we focused on detecting performance bugs that are important in multiple fields, such as mission critical applications, financial, and real time systems. Our software under study is a real time system used in the advertisement~/~marketing domain. 

Our research question was: \emph{How can we detect performance bugs using static code analysis and attributes extracted from source code repository?} To answer the question, we described and followed a methodological approach that resulted in application of four different prediction algorithms, namely: C4.5 Decision Trees, Na\"{\i}ve Bayes, Bayesian Networks, and Logistic Regression.

The best model (not taking into account information about file names\footnote{A novel project may not have sufficient amount of file-name-related data~\cite{fukushima2014esj}.}) is obtained with C4.5, using lines of code added, lines of code removed, file age, and file size as explanatory variables ($TPR = 0.73$, $ACC = 0.85$, and $PPV = 0.96$). The best model, which  has file name available to it, is based on Logistic Regression with all the attributes included: $TPR = 0.90$, $ACC = 0.91$, and $PPV = 0.92$.

Furthermore, the analysis of the variables, based on Logistic Regression model, can aid management in setting corrective actions. In particular, injection of performance bugs can be reduced by decreasing the amount of changes delivered on a single commit\footnote{This will lead to increase of the number of commits. However (based on the authors' industrial experience as well as the literature~\cite{porter1998, kemerer2009}) individual commits will become easier to review, leading to improved defect removal.}, as well as by focusing  quality assurance resources on a small subset of error-prone files (9\% of the total number of files).

We believe that our methodological approach is of interest to practitioners, because it provides them with a simple and tractable model, using easily extractable code attributes for predicting performance bugs on a new code commit. In our models we omit pattern type information to show that the data, extracted automatically from version control system, is sufficient to predict performance bugs. This suggests that performance bugs mapped to non-functional requirements can be detected using the same attributes (particularly ``lines added'') as functional bugs.  This simplifies adoption of the models, since complexity and lack of human and/or hardware resources are often discouraging reasons. It also speeds up detection of a performance bug, as the model is applied to new code before it was committed to public code branch. If the model detects the performance bug, developer decides whether to review the code deeply or proceed as-is.

The resulting factors affecting projects may vary from project to project; however, practitioners may replicate model creation process, described in this study, tailoring it to their needs. Moreover, we describe how to process an extremely imbalanced dataset (where the number of non-defective records is $>20$ times larger than the number of defective ones), which can help practitioners facing similar issues. 

The pattern identification that we conducted and briefly described in this paper, can also be considered as a guide for performance oriented best practices for the Python programming language. Additionally, these patterns can also be reused as static code analysis tools for identification of performance related fixes. They can also be considered as part of enhancement and development guidelines (e.g., Python's enhancement proposals \cite{pep}) for Python language.  Based on the methodology that we described, researchers can replicate similar pattern identification for other programming languages.

Our results are also of interest to theoreticians, since this work establishes a link between functional bugs and (non-functional) performance bugs, explicitly showing that attributes used for prediction of functional bugs can be used for prediction of performance bugs as well.

Going forward, we would like to extend our work to include other projects and company data.

\appendices

\section{Patterns}\label{app:patterns}

\subsection{Pattern 1}

An example of a Coding pattern is as follows. Python provides multiple methods to concatenate strings. For example,
% pattern #4
\begin{lstlisting}
    foo =  'a_string'
    bar = 'abc' + 'def' + str1
\end{lstlisting}
is much slower than
\begin{lstlisting}
    foo = 'a_string'
    bar = '{0}{1}{2}'.format('abc', 'def', str1)
\end{lstlisting}

\subsection{Pattern 2} \label{p:2}
An example of a Design pattern is as follows. Python provides a default logging mechanism, which, in some cases, cannot avoid expensive function calls and/or calculations, even if these are finally not necessary/logged. A simple workaround is by doing a cheap function call before executing the actual log function. Essentially, the ``cheating'' function is just making sure that the current line will be logged before the execution of the expensive call. For example,
% pattern #9
\begin{lstlisting}
    logger.setLevel(logging.CRITICAL)
    logger.debug(doSomethingExpensive())
\end{lstlisting}
the above example will spend all the time necessary to call the {\texttt{\small doSomethingExpensive()}} but finally won't use any of this information due to lower current logging level. However,  
\begin{lstlisting}
    logger.setLevel(logging.CRITICAL)
    if logger.isEnabledFor(logging.DEBUG):
        logger.debug(doSomethingExpensive())
\end{lstlisting}
preliminary check of the log level can prevent unnecessary calculations.
In case of the if-block returning \texttt{FALSE}, the expense of calling \texttt{\small doSomethingExpensive()} is saved. We made sure that the combination of an if-block and the call of the \texttt{\small isEnabledFor(<level>)} is always cheaper or, at the very least, the same as the default logging function. It might be worth mentioning that the application of the above pattern, resulted in a very significant ($>15\%$) performance improvement - however we understand that this might not apply to every system.

Illustration of the remaining nine patterns are given below.

\subsection{Pattern 3}
% pattern #1
Accessing global variables or built-in functions is slightly more time consuming than local. 
\begin{lstlisting}
    def slightly_slower(asequence, adict): 
        for x in asequence:
            adict[x] = hex(x)
\end{lstlisting}
\begin{lstlisting}
    def slightly_faster(asequence, adict): 
        myhex = hex
        for x in asequence: 
            adict[x] = myhex(x)
\end{lstlisting}
The latter is faster, because it is not accessing the built-in function globally in the second loop.

\subsection{Pattern 4}
% pattern #2
Addition is faster than multiplication.
\begin{lstlisting}
    x+x
\end{lstlisting}
is faster than 
\begin{lstlisting}
    x*2
\end{lstlisting}

\subsection{Pattern 5}
% pattern #3
\texttt{if} statements are expensive in Python. It is recommended to use \texttt{if} statements only if you know that they will be executed only once. If they are going to be used more than one time (e.g., if statement within a for loop) it may be efficient to replace them (if possible).

\subsection{Pattern 6}
% pattern #5
Loop unrolling - instead of a for loop:
\begin{lstlisting}
    def use_for():
        for i in range(1000):
            do(i)
\end{lstlisting}
use \texttt{map()}, if possible:
\begin{lstlisting}
    def use_map():
        map(do, range(1000))
\end{lstlisting}

\subsection{Pattern 7}
% pattern #6
Built in functions are the way to go in all cases because they are written in \texttt{C}, which makes them faster than any other logic you might use \cite{python_functions}.
However, there may also be differences in between them, e.g. \texttt{type()} vs. \texttt{isinstance()}, and \texttt{range()} vs. \texttt{xrange()} (which is resolved in Python 3).

\subsection{Pattern 8}
% pattern #7
Do not do:
\begin{lstlisting}
    for key in some_dict.keys()
\end{lstlisting}
but:
\begin{lstlisting}
    for key in some_dict
\end{lstlisting}

\subsection{Pattern 9}
% pattern #8
Cache a method called in a loop (this is easily applied on methods that return consistent results). Method look-ups can be expensive.
So for example in:
\begin{lstlisting}
    for i in xrange(1000):
        myobj.compute(i)
\end{lstlisting}
you can eliminate the look-up:
\begin{lstlisting}
    compute = myobj.compute 
    for i in xrange(1000):
        compute(i)
\end{lstlisting}

\subsection{Pattern 10}
% pattern #10
Swapping (memory optimization). Avoid:
\begin{lstlisting}
    temp = x 
    x = y
    y = temp
\end{lstlisting}
but do:
\begin{lstlisting}
    x, y = y, x
\end{lstlisting}

\subsection{Pattern 11}
% pattern #11
Try to do operations in place, instead of creating a new instance (memory optimization). Do not do:
\begin{lstlisting}
    sortme = sorted(sortme)
\end{lstlisting}
but instead:
\begin{lstlisting}
    sortme.sort()
\end{lstlisting}

\section*{Acknowledgments}
The authors would like to thank the RTB development team of Addictive Mobility for the invaluable discussions, suggestions and feedback they provided during this research study. The research is supported in part by MITACS Accelerate Grant \# IT04452 and NSERC Discovery Grant RGPIN-2015-06075.



\bibliographystyle{IEEEtran}
\bibliography{main_bib}

\end{document}